\documentclass[runningheads]{llncs}
\usepackage{graphicx}
\usepackage{xcolor}
\usepackage{enumitem}
\usepackage{subfiles}
\usepackage{listings}
\usepackage{xcolor}
\usepackage{listings}

\newcommand{\A}{A}
\newcommand{\B}{B}

\begin{document}
\title{Gimme That Model!: \\A Trusted ML Model Trading Protocol}
%
%
\author{Laia Amor\'{o}s \inst{1} \and
Syed Mahbub Hafiz \inst{2} \and
Keewoo Lee \thanks{Corresponding author} \inst{3} 
\and
M. Caner Tol \inst{4}}
\authorrunning{L. Amor\'{o}s et al.}
%
\institute{Aalto University, Finland\\
\email{laia.amoros@aalto.fi}
\and
Indiana University-Bloomington, USA\\
\email{shafiz@iu.edu}
\and
Seoul National University, Republic of Korea\\
\email{activecondor@snu.ac.kr}
\and
Worcester Polytechnic Institute, USA\\
\email{mtol@wpi.edu}}
\maketitle              
%

%
%
%
\section{Introduction}
Machine learning (ML) has achieved a tremendous success in making breakthroughs in various real-life problems from areas such as medicine, finances or social sciences, to mention a few. It has created a lucrative business model called Machine-Learning-as-a-Service (MLaaS), where big technological companies provide artificial intelligence (AI) services to customers. One of the functions of the MLaaS platform allows customers to purchase an ML model on demand. In the foreseeable future, we think that the market of trading ML models is going to grow significantly, and the security and privacy of the trade and the ML models will be crucial~\cite{technavio:mlaas:2019}. 

Let us illustrate this with an example. Suppose a car company $\B$ is interested in adding autonomous driving technology to its self-driving cars. $\B$ would like to buy a computer vision model, a pre-trained ML model, from an AI technological company $\A$. For instance, \textsf{Mobileye},
an advanced driver-assistance system (ADAS) provider, sells computer vision models installed on chips to automobile companies such as \textsf{Hyundai} or \textsf{Nissan}. In this scenario, $\B$ must have the ML model available without the need to go online: the system might lose the internet connection by accident or have no connection at all.

In another similar situation, imagine that a pharmaceutical company $\A$ has an ML model trained on the medical profiles of its patients. An IoT company $\B$ builds medical devices for its end-users. At some point, the end-users want to know whether they are prone to a specific disease through the medical device. $\B$ has neither the model nor the training data to provide such a service to its end-users, and it probably cannot share their private data due to privacy issues. $\B$  cannot use homomorphic encryption (HE) or multi-party computation (MPC) as privacy-preserving ML solutions offered previously because of computational resource requirements of the ML task.  Therefore, $\B$ might be interested in purchasing the ML model from $\A$ and install it on the medical devices, so that the end-users can check their medical status. 


During the trade of ML models, some particular interactions between two parties, e.g. a seller company $\A$ and a buyer company $\B$, need to be done before the final transaction. For instance, $\A$ wants to ensure that once the model is given to $\B$, $\B$ will buy it. In addition, $\B$ needs to confirm that the model is valid and accurate for its purposes before purchasing it. We propose an HE-based protocol, which will secure the transaction for both parties, allowing $\A$ to sell its ML algorithm to $\B$ keeping the model secret, and at the same time protecting $B$ from a possible \textit{bad} ML model before buying it.


There are two benchmarks of ML models to be tested before a transaction: accuracy and efficiency.
Whereas the timing result is independent of the input data (characteristic that ML models share with traditional software services), the accuracy of ML models depends excessively on training and test datasets. 
In this work we will focus on guaranteeing the accuracy of the ML models and leaving the efficiency issue on the background.

This report is organized as follows. In Section~\ref{s:noncryptosolutions}, we present the situation described above and propose some possible non-cryptographic solutions that turn out to be problematic. In Section~\ref{s:HE-sol}, we recommend our HE-based solution and describe possible improvements to the protocol to make the overall transaction more efficient and secure. In Section~\ref{s:discussion}, we start by discussing the feasibility of our HE-based solution, we then compare it to other potential existing cryptographic solutions, and we finish by considering a dual scenario: trading datasets instead of the ML models.

%



%

\section{Non-cryptographic Approaches and their Drawbacks}\label{s:noncryptosolutions}

In this section, we describe some non-cryptographic approaches (and their associated problems) to enable a company $\A$ to sell a pre-trained machine learning model to a customer $\B$. The situation is as follows: $\A$ is a company that owns a pre-trained ML model that a costumer $B$ would like to purchase. In this simple situation, on can think of (at least) three different approaches that do not make use of any cryptographic measures, and the consequent problems that might arise.

\begin{description}
    \item[Approach I:]\,The first approach is simple, but naive: $\B$ makes the payment before getting any service from $\A$. After the payment, $\A$ sends the pre-trained model to $\B$. The potential problem in this situation is that $\B$ has no idea about the model quality before the payment, as training and test accuracies or generalization ability cannot be tested beforehand. The model may not be powerful enough to deploy on the client's side due to poor hyperparameters or model selection. $\A$ may not even send a trained model at all.
    \medskip
    \item[Approach II:]\,In the second approach, when a client $\B$ wants to try the model, $\A$ sends the pre-trained model to $\B$. If the model fits well on $ \B$'s test dataset, $\B$ can make the payment for the model. The problem arising here is that $\B$ can deploy the model without making any payment, claiming that the model does not fit its needs, but keeping the model.
    \medskip
    \item[Approach III:]\,The third approach provides a test session before the contract to try to avoid the previous situations. Before any compromise between $\A$ and $\B$, $\B$ sends a test dataset to $\A$. $\A$ evaluates the trained model using $ \B$'s dataset and sends the results back. If the results are satisfying enough for $\B$, the payment is made to $\A$. The problem here is that $\B$ cannot know whether the predicted results are obtained from a well-trained ML model or, for instance, by hand. For example, a computer vision task may be manually achieved by crowdsourcing.
\end{description}

\section{Our HE-based Cryptographic Solution}\label{s:HE-sol}

In this section, we propose a potential solution for trading ML models that overcomes all the problems stated before. This solution is based on HE. After presenting the protocol, we continue by describing possible improvements to make the protocol more efficient. We then demonstrate some possible attacks to our protocol and suggest possible defenses. Finally, we summarize which ML model would be compatible with the proposed protocol while satisfying suitable efficiency and security.

\subsection{The Protocol}

\begin{figure}
\includegraphics[width=\textwidth]{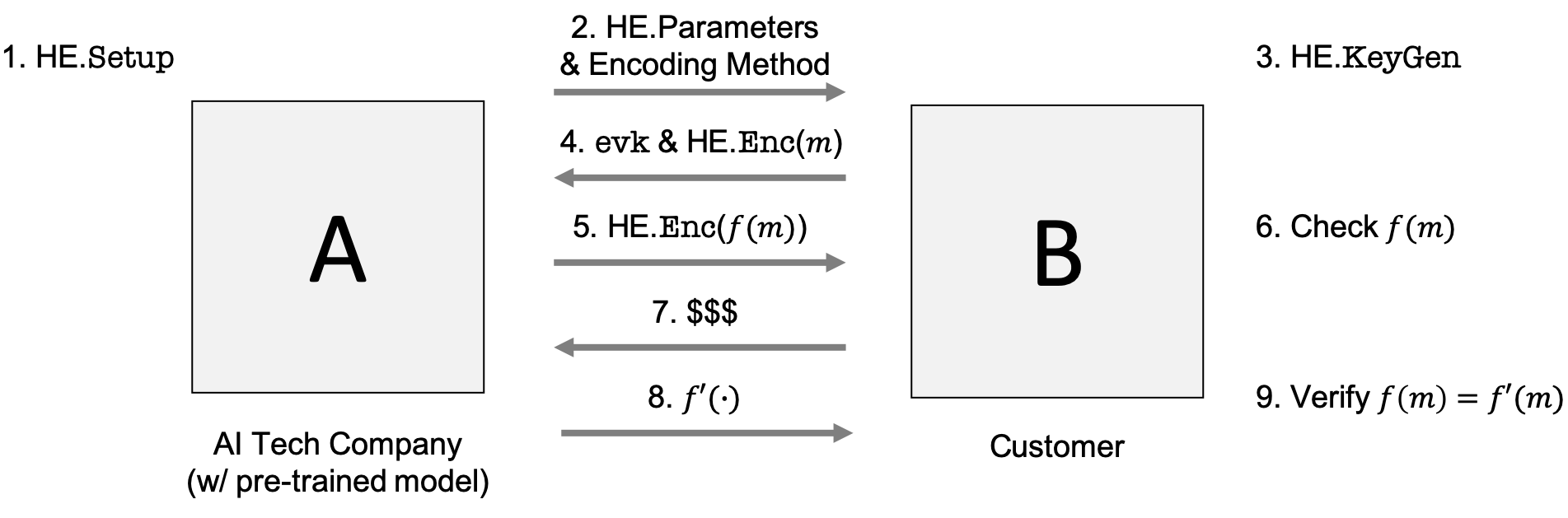}
\caption{Our HE-based Solution} \label{fig1}
\end{figure}

The steps of the proposed protocol are as follows.

\begin{enumerate}
    \item First $\A$ runs $\texttt{Setup}$ to obtain the public HE parameters (base cyclotomic ring, modulus, number of levels, etc.). Note that $\B$ cannot run $\texttt{Setup}$ if we use SHE since $\B$ does not know the model and the required depth accordingly. If we use FHE, $\B$ can run $\texttt{Setup}$.
    \footnote{SHE stands for \emph{somewhat} homomorphic encryption. FHE stands for \emph{fully} homomorphic encryption}
    
    \item $\A$ sends the HE parameters to $\B$. $\A$ also sends a data encoding method to $\B$. Agreement on the encoding method is essential since the encoding method considerably affects the performance, and $\A$ cannot manipulate the received ciphertexts easily.
    
    \item $\B$ runs $\texttt{KeyGen}$ after checking that the security level of the received HE parameters is sufficient.
    
    \item $\B$ sends the test data $m$ in encrypted form, i.e. sends $\texttt{Enc}(m)$, with evaluation keys $\texttt{evk}$.
    
    \item $\A$ performs homomorphic inference on $\texttt{Enc}(m)$ using the evaluation keys, i.e. computes $\texttt{Enc}(f(m))$, and sends it back to $\B$.
    
    \item $\B$ decrypts the received ciphertext and gets $f(m)$, and checks if the algorithm works as expected. 
    
    \item If $\B$ decides to buy the model $f(\cdot )$, $\B$ proceeds to do the payment to $\A$.
    
    \item $\A$ sends the model $f'(\cdot)$ to $\B$.
    \item $\B$ checks if $f(m)$ is equal to $f'(m)$. If not, $\B$ now has the evidence that $\A$ did not send the proper model. 
\end{enumerate}

Our HE-based protocol shows substantial advantages over the other non-cryptographic approaches presented in Section~\ref{s:noncryptosolutions}. As the protocol is built on \emph{homomorphic} encryption, $\B$ can have a test session without even knowing $\A$’s model before the contract. 
The \emph{security} of the HE scheme ensures that $\A$ cannot cheat during the test session as in the non-cryptographic Approach III, because $\A$ does not see the test data $m$. The received message $f(m)$ in step 6 ensures the \emph{commitment} by allowing $\B$ to notice in step 9 if $\A$ cheats by not sending a proper model in step 8.

\subsection{Efficiency of the Protocol}
In general, the main drawback of HE in practical usage is its slow speed. One attractive feature of our solution as an application of HE is that, in our setting, latency issues are more acceptable, since no real-time computation is necessary. Depending on the importance of the ML model transaction, a seller and a buyer may take their time for securing the trade. However, requiring an excessive amount of inference time would cause inconvenience. In this section, we discuss possible optimizations for our solution.

Optimization methods for general HE applications are also applicable to our solution. For example, a buyer can batch several test data into a ciphertext, and then the seller can evaluate the ML model only once for multiple test data. One would probably want to use the CKKS scheme for approximate homomorphic computations, and approximate the ML model into an HE-friendly version, i.e. into multivariate polynomials. Since these may cause other errors to the output, one should carefully consider this trade-off between accuracy and efficiency. In this case, the buyer should keep in mind that a new glitch has occurred, and the original model will be slightly more precise than the homomorphically evaluated results. 
However, since many ML algorithms are robust to errors in general, this might not cause a big problem.

Another possible optimization is to use HE parameters with a reduced security level, rather than HE parameters with conventional security level (e.g., 128-bit security). Since we need the security of ciphertexts only until the agreement of the contract is made, we can use more efficient parameters.

\subsection{Towards the Perfect Model Protection}

Even though HE provides strong security in our scenario, there are still some remaining issues that need to be considered due to the nature of the ML model trading situation and the lack of circuit privacy in concurrent HE schemes.
In this section, we consider possible attacks and suggest defenses for these attacks.

To begin with, a malicious buyer may perform numerous test queries to obtain nontrivial information about the ML model (e.g., a model extraction attack~\cite{TramZJRR:2016}).
To prevent a model extraction attack, the seller must limit the number of trials that a buyer can query. This can be done both explicitly and implicitly. Explicitly, a seller can regulate the amount of queries.
Implicitly, a seller can require costs to a buyer for each query. This cost works as a client puzzle: for a small number of queries, the prices are negligible; but for a large number of queries to perform the model extraction attack, the costs are infeasible to pay. 

Another natural concern on the ML model trading scenario is malicious redistribution. That is, once an ML model is sold, the buyer can illegally resell or redistribute it. In this respect, a possible solution is to use \textit{neural network watermarking}~\cite{AdiBCPK:2018,RouhaniCF:2019}. 

A third concern is the circuit privacy of HE schemes: since concurrent HE schemes do not provide circuit privacy, there exists a possibility of information leakage of ML models from the output of homomorphic computations. However, there are well-known countermeasures such as noise-flooding or bootstrapping~\cite{BoursePMW:2016}.

\subsection{Compatible ML models}

In terms of security and efficiency, the following properties of an ML model need to be taken into account in order to work with the proposed protocol.

\begin{description}
    \item[Criteria 1:]\,The ML model needs to be complex enough so that it can resist a model extraction attack.
    \item[Criteria 2:]\, The homomorphically encrypted ML model should have plausible inference time to evaluate the test dataset.
\end{description}

For binary and multi-class logistic regression models, the size of the feature vector should be much greater than \( k+1\) and \(c\,(k+1) \) respectively, where \(k\) is  the number of allowed queries and \(c\) is the number of classes. Neural networks like DNN, CNN, RNN, etc., should have network parameters much larger than \(k\) (e.g., more than \(100\,k \) regarding the experimental results of \cite{TramZJRR:2016}).

\section{Discussions}\label{s:discussion}

In this section, we first discuss the plausibility of our trading ML models scenario. We then consider possible alternative cryptographic solutions to the problem like multiparty computation or zero-knowledge arguments. We end the section by considering a dual situation: trading datasets instead of ML models.

\subsection{Plausibility of Trading ML Models}
Machine learning-based solutions have become a rather conventional way of tackling many problems. However, training ML models requires a substantial amount of computational power and a vast amount of training data. Therefore, selling such trained ML models can be seen as a rosy business model, since MLaaS market is growing fast. Many companies in this area propose to train and tune the models of a customer at a reasonable price (compared to the cost of computing resources and massive datasets required if the customer were to train the model by itself). 

One may ask why the company should sell the model itself instead of providing online access and sell subscriptions. We are going to answer this question from various perspectives. {First of all}, there are several ML applications where the service needs to work also when the system goes offline (e.g., driverless car). Another disadvantage is the privacy issue. A customer has to send always its private data to the service provider if this only offers a subscription. Of course, we can use HE or MPC to solve this privacy issue via PPML. However, this PPML approach suffers from slow speed and 
there are some ML applications where such latency cannot be tolerated (e.g., driverless car).

Another point is that a customer might want to train the model further to improve its performance for new data points using online learning, or the model can be used in multitask learning~\cite{Caruana:1997}. 
Without trading the model itself, this would not be possible for a customer. We also note that trading ML models is a one-time sale, but the company can retrain the model with up-to-date datasets after a while and sell updates.

\subsection{Alternative Cryptographic Solutions}

Secure multiparty computations (MPC) can be used as an alternative solution to our HE-based approach for the safe ML model trading scenario. When using MPC, one can hide the model weights, but it is a non-trivial task to protect the whole ML model, including its hyperparameters. That is, a seller and a buyer would share the ML model structure and perform secure multiparty computations on it, without revealing the model weights and the test data to each other. This leads to the solution being more susceptible to model extraction attacks. Moreover, the selection of suitable neural network architectures for a given problem and dataset is often difficult and resource-consuming~\cite{msrblog:petridish:2019}, meaning that sharing the structure of the ML model might already be a substantial loss of intellectual property for the seller.

Another issue for an MPC-based solution is communication overhead. The communication cost between the buyer and the seller becomes more significant as the ML model becomes more complex, and the two parties should remain online during the computation. This might be a minor problem if two large companies do the transaction, but might be an essential issue if a customer is an individual with low computing power. 

One might also consider zero-knowledge arguments (e.g., zk-SNARK) for another possible solution to the model trading scenario. In particular, when test data is given, the seller can use zk-SNARK to prove that it has the ML model, which outputs one specific result without revealing the parameters (zero-knowledge). However, in this case, the buyer has to disclose its test data to the seller, leading to the same problems as the non-cryptographic Approach III described in Section~\ref{s:noncryptosolutions}.

\subsection{Dual Scenario: Trading Datasets}

\begin{figure}
\includegraphics[width=\textwidth]{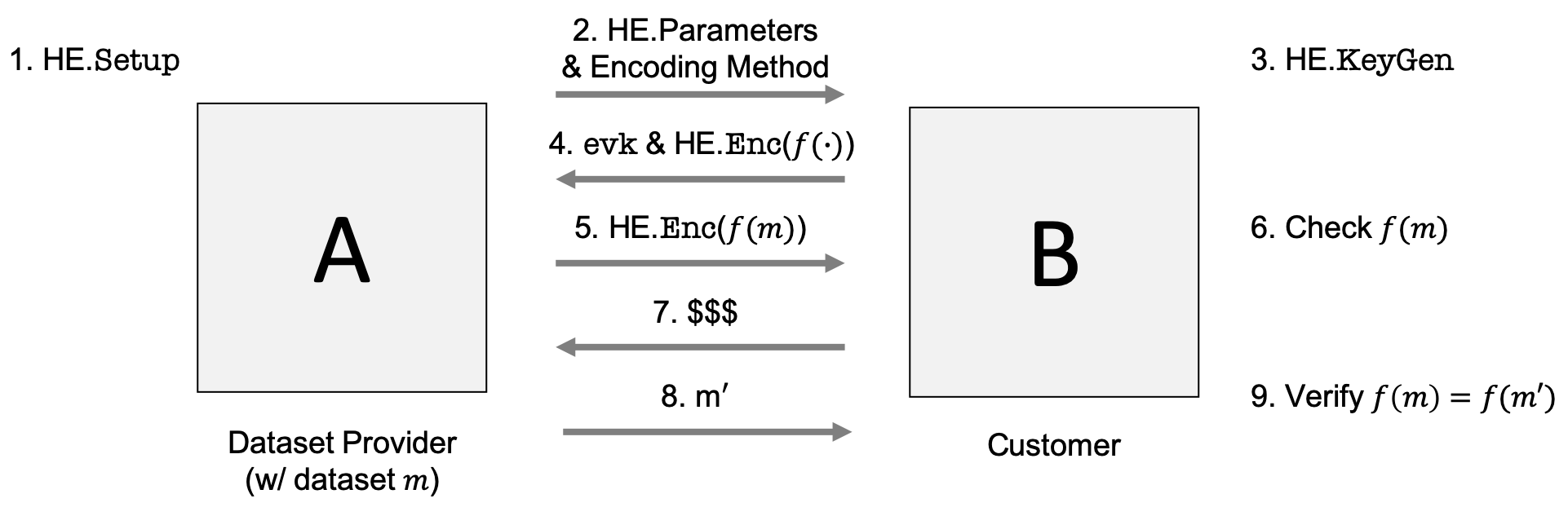}
\caption{Our HE-based solution for the dual scenario.} \label{fig2}
\end{figure}

One interesting point of view about our solution is that one can also consider its \emph{dual} scenario. In the original situation, the seller has a function (e.g. a ML model), and the buyer wants to check the output of the function on particular data before the transaction. On the other hand, in the dual scenario, the seller has a dataset, and the buyer wants to check the output of a particular function on the dataset before the transaction. Thus, the roles of functions and data are switched. We can directly apply our protocol, which is designed for the original scenario, to the dual situation by just interchanging the roles of function and data. 
Our HE-based solution for the dual scenario is summarized in Figure~\ref{fig2}.

However, in this case, our solution seems to have a less significant advantage over an existing MPC-based solution~\cite{GiladLLRR:2019}.
This is because, unlike the original solution, an ML model is being sent over a homomorphically encrypted channel. In this scenario, we probably want to encrypt only the model weights and biases and send the model hyperparameters unencrypted. This invalidates the first advantage of the HE-based solution over the MPC-based solution, which is described above. If not, we can homomorphically encrypt the whole circuit description. However, this approach does not look practically sound again.

\bibliographystyle{plain}
\bibliography{references}

\begin{thebibliography}{1}

\bibitem{AdiBCPK:2018}
Yossi Adi, Carsten Baum, Moustapha Cisse, Benny Pinkas, and Joseph Keshet.
\newblock Turning your weakness into a strength: Watermarking deep neural
  networks by backdooring.
\newblock In {\em Proceedings of the 27th USENIX Conference on Security
  Symposium}, SEC'18, pages 1615--1631, USA, 2018. USENIX Association.

\bibitem{BoursePMW:2016}
Florian Bourse, Rafa\"{e}l Pino, Michele Minelli, and Hoeteck Wee.
\newblock Fhe circuit privacy almost for free.
\newblock In {\em Proceedings, Part II, of the 36th Annual International
  Cryptology Conference on Advances in Cryptology --- CRYPTO 2016 - Volume
  9815}, pages 62--89, Berlin, Heidelberg, 2016. Springer-Verlag.

\bibitem{Caruana:1997}
Rich Caruana.
\newblock Multitask learning.
\newblock {\em Machine Learning}, 28(1):41--75, Jul 1997.

\bibitem{RouhaniCF:2019}
Bita Darvish~Rouhani, Huili Chen, and Farinaz Koushanfar.
\newblock Deepsigns: An end-to-end watermarking framework for ownership
  protection of deep neural networks.
\newblock In {\em Proceedings of the Twenty-Fourth International Conference on
  Architectural Support for Programming Languages and Operating Systems},
  ASPLOS'19, pages 485--497, New York, NY, USA, 2019. Association for Computing
  Machinery.

\bibitem{msrblog:petridish:2019}
Debadeepta Dey.
\newblock Microsoft research blog: Project petridish: Efficient forward neural
  architecture search, 2019.

\bibitem{GiladLLRR:2019}
Ran Gilad-Bachrach, Kim Laine, Kristin Lauter, Peter Rindal, and Mike Rosulek.
\newblock Secure data exchange: A marketplace in the cloud.
\newblock In {\em Proceedings of the 2019 ACM SIGSAC Conference on Cloud
  Computing Security Workshop}, CCSW'19, pages 117--128, New York, NY, USA,
  2019. Association for Computing Machinery.

\bibitem{technavio:mlaas:2019}
technavio.
\newblock Global machine learning-as-a-service (mlaas) market 2019-2023, 2019.

\bibitem{TramZJRR:2016}
Florian Tram\`{e}r, Fan Zhang, Ari Juels, Michael~K. Reiter, and Thomas
  Ristenpart.
\newblock Stealing machine learning models via prediction apis.
\newblock In {\em Proceedings of the 25th USENIX Conference on Security
  Symposium}, SEC'16, pages 601--618, USA, 2016. USENIX Association.

\end{thebibliography}
\end{document}